# Tap tactile presentation by airborne ultrasound

Haruka Danil Tsuchiya[1], Zen Somei[1], Yasutoshi Makino [1] and Hiroyuki Shinoda[1]

[1] *Graduate School of Frontier Sciences, The University of Tokyo, Japan*

(Email: tsuchiya@hapis.k.u-tokyo.ac.jp)

**Abstract ---** **The airborne ultrasound tactile display can present tactile information without direct contact. Using this technology, we developed two methods for simulating the tactile sensation of tapping an object with a finger: the Amplitude Modulation Method and the Lateral Modulation Method. The first method, Amplitude Modulation, simulates the tactile sensation of tapping a soft, deformable surface, like a deflated balloon. The second method, Lateral Modulation, simulates the tactile sensation of a rigid surface that easily resonates with vibrations, like a cymbal. In the demonstration, participants can experience the difference between these two tactile stimuli by tapping virtual objects displayed on a screen.**

## 1 INTRODUCTION

Tap tactile sensation is a fundamental tactile element that always occurs at the moment of contact with an object. This perception is instantaneous in terms of time length, but there is a tactile range within it, which varies depending on the texture of the object being contacted. Reproducing this is important for a more natural tactile representation of object contact.

Airborne ultrasound haptic displays are expected to provide a high level of immersion because they can present tactile sensations without physical contact with the device, thus not restricting user movements during VR experiences. Recently, in addition to vibration sensations, the ability to present pressure sensations has been achieved, leading to research on rendering various tactile sensations using airborne ultrasound. In this study, we tackled the challenge of how well the sensation of tapping an object surface with a finger can be represented using airborne ultrasound. We demonstrated the potential to apply this method as a new haptic feedback technique to improve the operability of airborne interfaces and to present different material textures.

## 2 PROTOTYPE

### 2.1 Airborne Ultrasound Tactile Display

In this study, we use an Airborne Ultrasound Phased Array device (AUTD) to present tactile sensations [1]. The AUTD emits ultrasound waves and provides tactile feedback at the focal point using the radiation pressure generated by the focused ultrasound. To enhance the tactile stimulation provided by airborne ultrasound to a level perceivable by humans, modulation of the ultrasound is necessary. There are two modulation methods: Amplitude Modulation (AM) and Lateral Modulation (LM) [2]. The AM method presents vibratory tactile stimuli by periodically varying the amplitude of the radiation pressure of the ultrasound. On the other hand, the LM method can present not only stronger vibratory sensations than the AM method but also pressure tactile stimuli by periodically varying the focal position of the ultrasound. In this study, we primarily use LM-based tactile stimulation.

### 2.2 Proposed Methods

The results of organizing the tactile components when tapping an object with a finger are shown in Figure 1. First, the initial moment of contact with the object produces the strongest impact sensation, and this vibratory impact rapidly diminishes to a level where it is no longer perceptible. We define the period during which this sensation can be perceived as the "attenuation collision phase." Subsequently, if the contact with the object is maintained, only a constant normal force from the object is perceived. We define the period during which this sensation can be perceived as the "stationary phase."

In expressing the perception process of the tapping sensation through haptics, we identified two necessary types of tactile stimulation. The first is the stimulation of "gradually weakening damped vibrations" in the attenuation collision phase, and the second is the stimulation of "a constant pressure sensation" in the stationary phase. To reproduce these sensations, as shown in Figure 2, we developed two methods for the attenuation collision phase and one method for the stationary phase.

**Amplitude Modulation Method**: This method involves varying the amplitude intensity of the ultrasound. By gradually reducing the amplitude fluctuation of the radiation pressure and increasing the amplitude strength to its maximum value, it is possible to create a stimulation where the vibratory sensation diminishes.

**Lateral Modulation Method**: This method involves varying the distance of the focal point of the ultrasound, referred to as "size." By decreasing the size from initially large to progressively smaller, it is possible to create a stimulation where the vibratory sensation weakens.

For the stimulation in the stationary phase, we used a technique developed in prior research [3]. Specifically, by controlling the LM modulation frequency between 5-15 Hz and adjusting the size to below 1 mm, we were able to produce pressure sensation with minimal vibratory sensation.

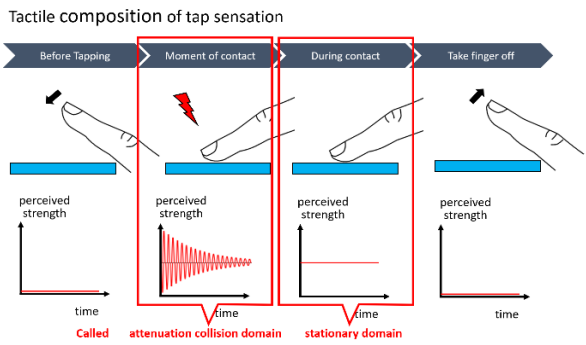

Fig.1  Tap composition of tap tactile sensation

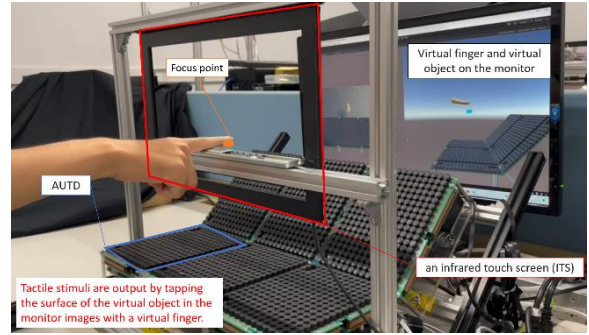

Fig.2  Proposed 2 tactile stimulation methods of tap sensation

### 2.3  System Configuration

The system configuration is shown in Figure 3. The device consists of six AUTDs and an infrared scanning touch panel (ITS), all controlled by Unity. This setup allows for real-time acquisition of the 2D coordinates of the finger, enabling control of a virtual finger on the display. The monitor shows both the virtual finger and the virtual object, and tactile feedback is provided when the virtual finger on the screen touches the surface of the virtual object. With this system, users can experience the tactile sensation of tapping by interacting with the surface of a virtual object using their finger.

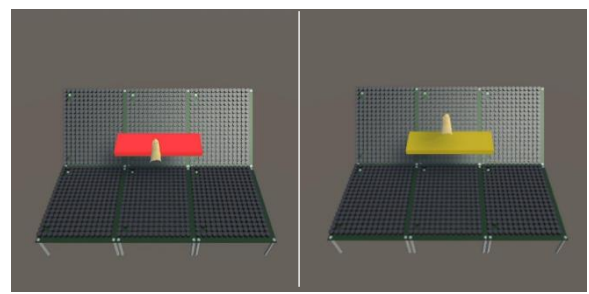

Fig.3  System configuration

### 2.4  Demonstration

Fig. 4 shows the demonstration setup. When tapping the red object, users experience a tactile sensation like tapping a deflated balloon. When tapping the yellow object, users experience a tactile sensation like tapping cymbals. Specifically, the former provides a tactile sensation of a soft, elastic object with a deformed surface, akin to a deflated balloon, by using Amplitude Modulation in the attenuation collision phase. The latter provides a tactile sensation of an object with a non-deformed surface but with easily resonating vibrations, like cymbals, by using Lateral Modulation in the attenuation collision phase.

Fig.4  User's point of view during the demonstration: The red object represents the tap tactile sensation of a deflated balloon, while the yellow object represents the tap tactile sensation of cymbals

### 3  CONCLUSION

We reproduced the tactile sensation of tapping, which consists of attenuated vibratory stimulation in the attenuation collision phase and pressure sensation in the stationary phase, using airborne ultrasound. In this process, we proposed two methods developed for attenuated vibratory stimulation: the Amplitude Modulation Method and the Lateral Modulation Method. The demonstration showed that the Amplitude Modulation Method could represent objects with soft elastic forces, like a deflated balloon, while the Lateral Modulation Method could represent objects that resonate with vibrations, such as a cymbal.